\documentclass{kapproc} 

\def\etal{{\it et al.\ }}
\def\ie{{\it i.e.\ }}
\def\ni{{\noindent}}
\def\eg{{\it e.g.\ }}
\def\th{\thinspace}
\def\approxgt{\,\raise2pt \hbox{$>$}\kern-8pt\lower2.pt\hbox{$\sim$}\,}
\def\approxlt{\,\raise2pt \hbox{$<$}\kern-8pt\lower2.pt\hbox{$\sim$}\,}
\long\def\jumpover#1{{}}

\usepackage{graphicx}

\begin{document}

\articletitle{Nonlinear Properties of the\\
 Semiregular Variable Stars}

\author{J. Robert Buchler}
\affil{University of Florida, USA}
\email{buchler@phys.ufl.edu}
\author{Z. Koll\'ath}
\affil{Konkoly Observatory, Budapest, HUNGARY}
\email{kollath@konkoly.hu}
\begin{abstract}
We demonstrate how, with a purely empirical analysis of the irregular 
lightcurve data, one can extract a great deal of information about the stellar 
pulsation mechanism.
An application to R~Sct thus 
shows that the irregular lightcurve is the result of
the nonlinear interaction of two highly nonadiabatic pulsation modes, 
namely a linearly
unstable, low frequency mode, and the second mode that, although linearly 
stable, gets entrained through a 2:1 resonance. 
In the parlance of nonlinear dynamics the pulsation is the result of a 4
dimensional chaotic dynamics.
\end{abstract}

\begin{keywords}
Stellar Pulsations, Variable Stars, Chaos
\end{keywords}

In the following we lump together under the label 'semiregular' {\it largo
sensu}, the stars of RV Tau type, the Semi-Regular stars and some of the Mira
variables.  All of these stars have lightcurves of varying degrees of
irregularity.  Most of our information about them comes from amateur astronomer
data bases (AAVSO, AFOEV, BAAVSS and VSOLJ) which contain data on a large
number of bright stars spanning almost a century.  Many have at least several
decades of good temporal sampling.  Unfortunately, the data are very noisy,
especially around the lightcurve minima, and for our analyses we therefore have
to bin, smooth and filter the data to form a time-series with equal
time-intervals.  Figures 1 and 2, on top, show sections of lightcurves for 4
selected stars, {\it viz.} R~Sct (RV~Tau type), R~UMi (SR type), R~Cyg and
X~Aur (both Mira type).

\begin{figure*}[fig1]
\centerline{\resizebox{11.8cm}{!}{\includegraphics{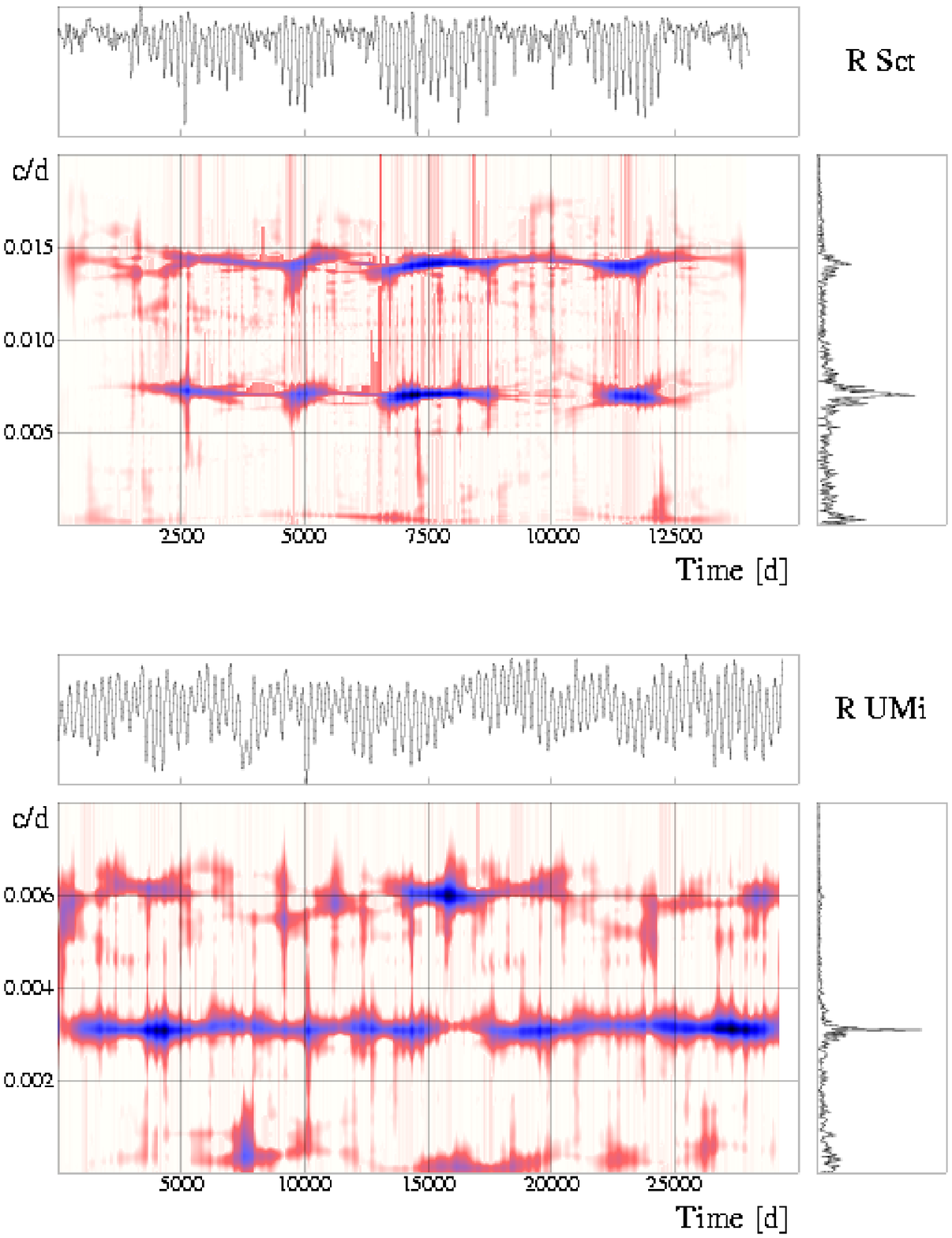}}}
\caption{Time-frequency plots for R~Sct and R~UMi; see text for details.}
\end{figure*}

\begin{figure*}[fig2]
\centerline{\resizebox{11.8cm}{!}{\includegraphics{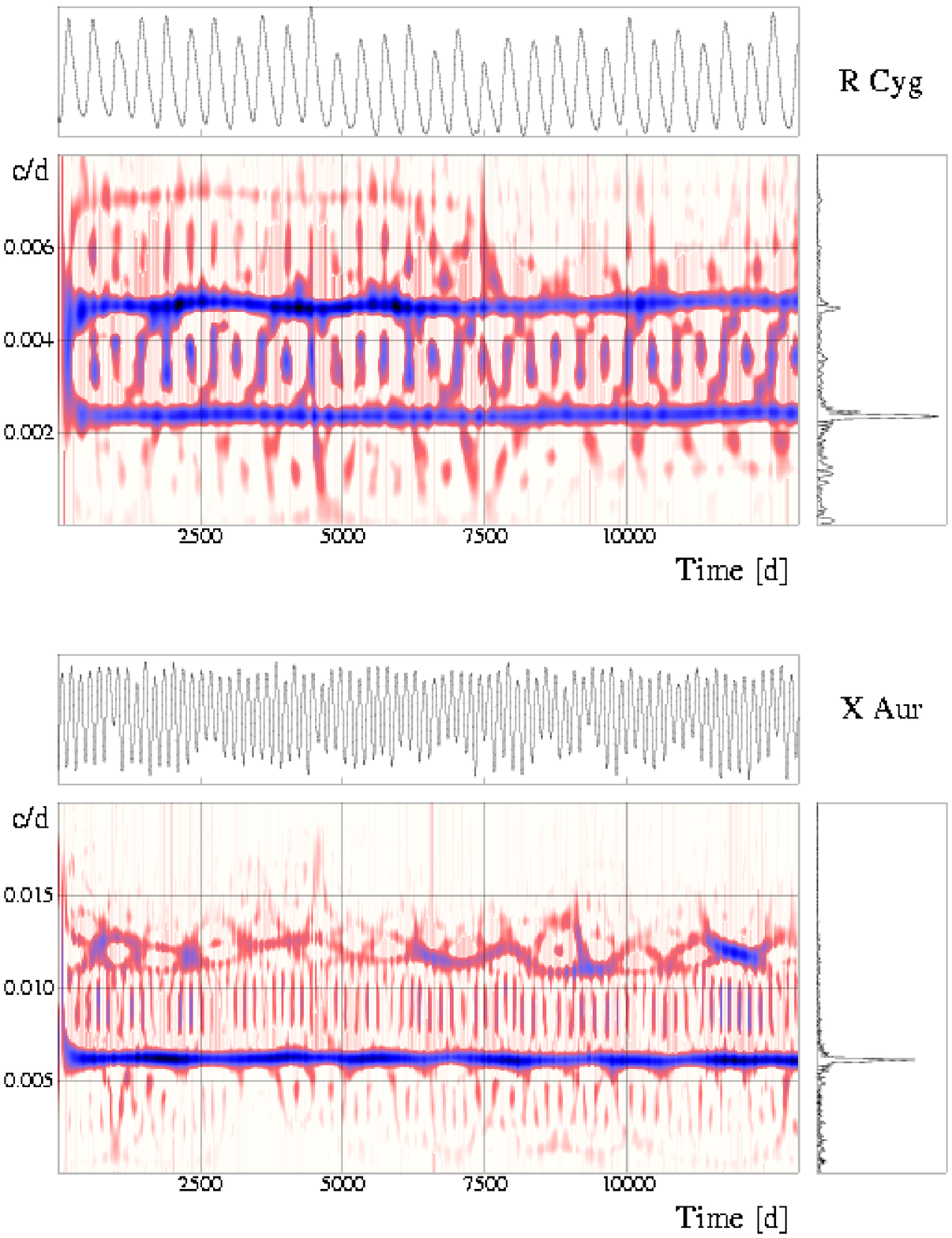}}}
\caption{Time-frequency plots for R~Cyg and X~Aur.}
\end{figure*}

The questions that we are asking here are: What are the nature and cause of the
irregularities?  Is there one or several underlying mechanisms?  Do the the
lightcurve data contain any quantitative information that can be extracted and
exploited?  If so, how does it correlate with luminosity, and can it be used
for distance measurements?  Most past work ({\it cf.} these conference
proceedings) has concentrated on extracting period-luminosity relations which
sheds light on the evolutionary and pulsational status and on the stellar
structure of these stars.  But is there additional information in the data?

Figures 1 and 2, on the right side, display the Fourier spectra of the 4 sample
stars.  One might be tempted to classify R~UMi and X~Aur as monoperiodic and
R~Sct and R~Cyg as biperiodic as in Kiss \etal (2002).  However, these stars
are {\sl not multiperiodic}.  Neither the amplitudes, nor the phases, nor even
the frequencies are constant in time.  This can best be seen in the
corresponding time-frequency plots.  Instead of the more common wavelet or
Gabor transform we have made our time-frequency plots with a Choi-Williams
kernel (Cohen 1994) which has the property of sharpening features ({\it cf.}
also Koll\'ath \& Buchler 1997).  It is not astonishing that the {\sl
instantaneous amplitude} in the dominant peak ($f_0$) is seen to vary (dark
corresponding to higher values on the adopted greyscale), but very
interestingly the {\sl instantaneous frequency} varies as well!  In order to
make the structure of the 'harmonic' region ($\approx 2f_0$) visible on the
same greyscale we have scaled up the amplitudes in that region.  Remarkably,
the harmonic frequency does not move synchronously with the dominant frequency.
Furthermore, for R~Sct, the $2f_0$ power seems to switch back and forth between
$2 f_0$ and $2f_0 + \Delta f$.  A similar behavior occurs in X~Aur.

None of these features of the time-frequency plots, nor the irregularity of the
lightcurves for that matter, can be explained by evolution, by dust or spots,
by binarity or by stochasticity, even though all of these effects can be
present and influence the lightcurve.  Instead, the time-frequency plots
suggest a low dimensional underlying chaotic dynamics consisting of the
nonlinear interaction of a few modes.  We suggest that multimode might be a
better label for these stars than multiperiodic.

In those cases where the amateur data and the Cadmus (private communication,
and Buchler \etal 2002) data overlap, the time-frequency analysis gives
essentially the same results - \eg these fingerprints of nonlinear mode
interactions are insensitive to observational noise.

\vskip 10pt

We wish to stress that our nonlinear approach to the study of the pulsations of
the semiregular stars, namely the {\sl global flow reconstruction} is fully
{\em empirical}, \ie devoid of theoretical modelling.  Only two working
assumptions are made:

1. {\sl the lightcurve is produced by the
(deterministic) nonlinear interaction of a small number of pulsation modes},

2. {\sl the system is autonomous}, \ie we ignore time dependence such as
evolution over the span of the data.

\ni For additional details we refer to Buchler \& Koll\'ath (2000) and Buchler 
\etal (1996). 
Our assumptions imply that the star's behavior is describable by 
a differential system in a \textbf{physical phase space} 
of {\it a priori} unknown dimension $d$.
\begin{equation}
{d {\bf x}\over dt} = {\bf F}({\bf x})
\label{phys}
\end{equation}
where ${\bf x}$ is a $d$-dimensional vector whose components are the phasespace
variables (which could be modal amplitudes and phases, for example).  
For a single oscillatory 
mode, \eg $d$ would be equal to 2, for 2 coupled oscillatory modes $d=4$.
The involvement of a  secular mode would add 1 to the dimension.

In parallel we now introduce a \textbf{reconstruction space} of dimension $d_e$,
in which we construct successive position vectors 
$${\bf X^n} = (s_n, s_{n-\tau}, s_{n-2\tau}, \ldots, s_{n-(d_e-1) \tau})\th ,$$
using the observational
data $s_n = s(t_n)$, the magnitude in our case.  The quantity $\tau$ is called
the delay parameter.  If our assumptions are satisfied, then the temporal
behavior should be captured by an evolution equation 
\begin{equation} 
{\bf X}^{n+1} = {\bf \cal M}({\bf X}^n) \th,
\label{map} 
\end{equation}
provided $d_e$ is large enough.  We could also have introduced a differential
system akin to \ref{phys} in this space -- our ${\bf X}^n$ is merely a
stroboscopic description of the dynamics.  The map ${\bf\cal M}$ is assumed to
be a sum of all the multivariate monomials up to some order (usually 4) and the
unknown coefficients are determined by a least squares fit from the data.\th (We
minimize $||{\bf X}^{n+1} - {\bf \cal M}({\bf X}^n)||$ over the data set).  A
powerful embedding theorem assures us that the dynamics in the physical
phasespace and in the reconstruction space are the same provided that $d_e$ is
large enough.  Consequently, {\sl from the study the behavior of the
reconstructed system we can infer otherwise unknown properties of the
physical phase space.}  

Once we have constructed a map from the data set we can iterate that map and
generate '{\sl synthetic signals}' which are much longer than the observational
data set.  From the latter we can then compute Lyapunov exponents and the
fractal dimension $d_L$.  We consider a reconstruction successful when the
synthetic signals are robust with respect to a range of smoothing parameters
and a range of delay parameters $\tau$, are stable, and when the results are
independent of the embedding dimension $d_e$, as long as the latter is large
enough.  The lowest value for which we obtain robust results will be called
$d_e^{min}$.

\vskip 10pt

In Buchler \etal (1996) we analyzed the AAVSO data of R~Sct.  Here we have
repeated our analysis with a richer data base obtained by combining the AAVSO,
AFOEV, BAAVSS and VSOLJ data and extending the basis to date.  In Fig.~4 we
display the R~Sct lightcurve together with a typical synthetic signal in 4D.
The synthetic signal is clearly seen to capture the nature of the lightcurve.
This becomes even more evident when one looks at the lightcurve data over the
last 150 years (Koll\'ath 1990).  With the extended data basis our results
do not change and remain very interesting.

1. In 3D no robust reconstructions are possible.  The results are not changed
   by going to $d_e$ = 5 and 6.  They are also independent of the delay
   parameter $\tau$ within a broad range.  We conclude that $d_e^{min} = 4$.

2. One of the Lyapunov exponents is always positive, implying that the
   pulsation is chaotic.  The fractal dimension of the attractor is $d_L \sim
   3.2$.  The values of the exponents and of $d_L$ is largely independent of
   $d_e$.

3. Clearly the (Euclidean) dimension of the physical phasespace is sandwiched
   between $d_L=3.2$ and $d_e^{min}=4$.  {\sl We therefore can infer that
   $d=4$}. This suggests that the lightcurve is generated by the nonlinear
   interaction of two vibrational modes, consistently with the time-frequency
   analysis.

4. When the map is linearized around its fixed point
one obtains 2 spiral stability roots\hfill\break
$\lambda_1 = \pm i \th    0.0068 \times 2\pi \thinspace + 0.0044 d^{-1}$, and
$\lambda_2 = \pm i \th    0.0145 \times 2\pi \thinspace - 0.0062 d^{-1} .$
\hfill\break
Because the fixed point of the map corresponds to the equilibrium state of the
star, the two spiral roots corroborate the presence and excitation 
of two vibrational modes.  Furthermore it tells
us that there is a first, {\sl linearly unstable mode} of frequency  
$f_0 = 0.0068$ and a second mode, {\sl linearly stable} one, 
with frequency slightly 
greater than $2 f_0$.  This is in agreement with the time-frequency analysis.
Note that these modal properties come from a map which was obtained through a
   fit to the data.   None of these properties were imposed.

\begin{figure*}[fig3]
\centerline{\resizebox{11.8cm}{!}{\includegraphics{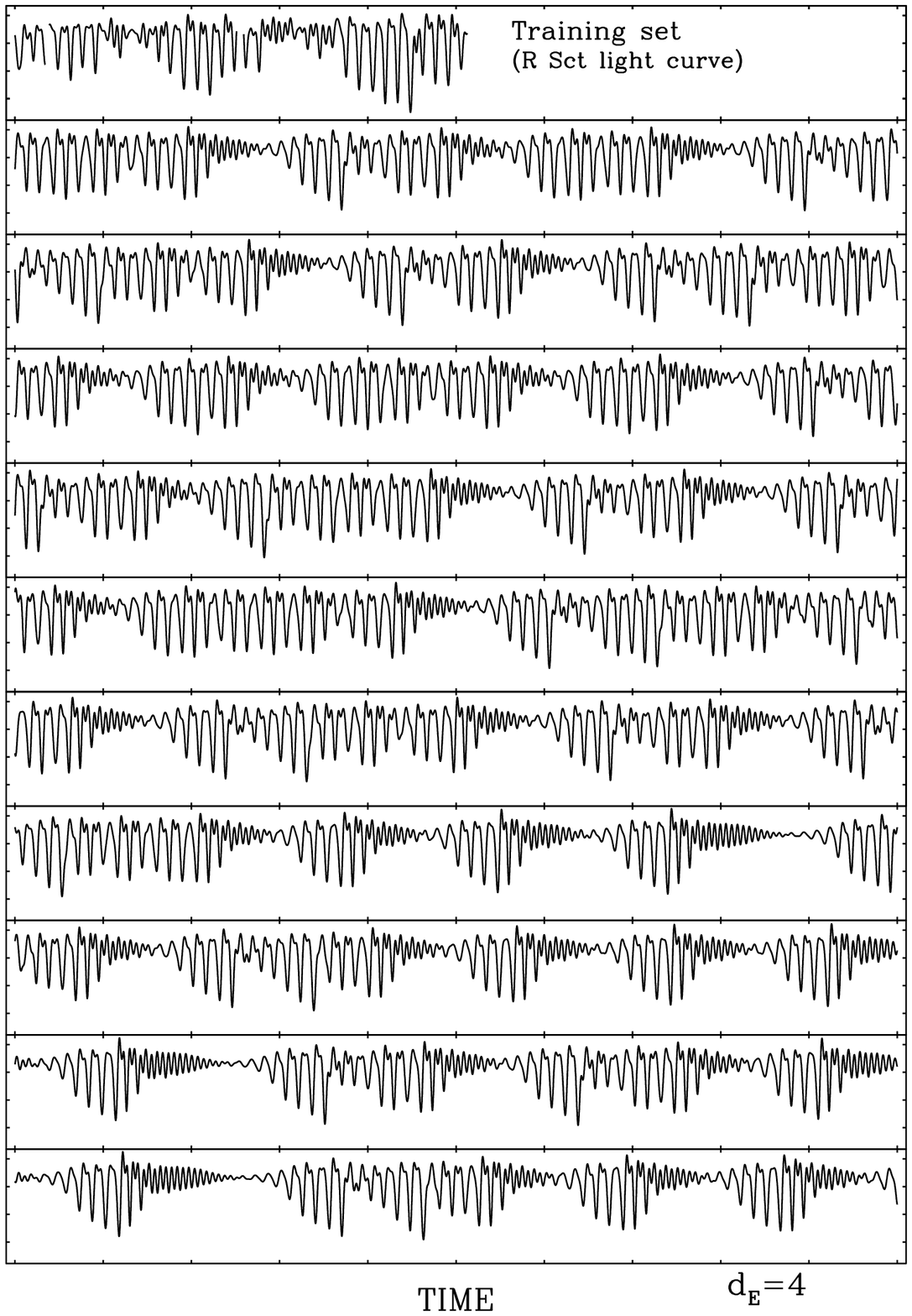}}}
\caption{R~Sct: {\sl top:} section of R Sct light curve; {\sl below:} long
synthetic signal generated with a 4D map.}
\end{figure*}

\vskip 10pt

We conclude that there is no need for a {\it deus ex machina}, such as
irregular convective overshoots, to explain the nature of the irregular
pulsations.  From our empirical data analysis we have arrived at a useful
physical picture: {\sl The irregular pulsation is the result of the nonlinear
interaction of two strongly nonadiabatic pulsation modes}.  A lower frequency,
linearly unstable mode entrains a stable, higher frequency one through a 2:1
resonance.  The unstable mode wants to grow, but shares kinetic energy with the
stable one which then dissipates it, and the cycle repeats.  Because of the
strongly nonadiabatic nature of these modes the motion is irregular (chaotic).

The reader may wonder why the same resonant scenario gives rise to a
synchronized, periodic pulsation in the classical bump Cepheids (Buchler
1993). The physical reason is that the latter are only weakly nonadiabatic, the
ratio $\eta$ of growth rate to frequency is of the order of 0.01.  In the
semiregulars the ratio of luminosity to mass is more than ten times larger,
and $\eta$ is of order unity.  The fact that the amplitude can vary on
the timescale of the period is of course a necessary (but not sufficient)
condition for chaotic behavior.

The amateur astronomer data bases contain a number of semiregulars with
sufficient coverage to allow the same approach.  We have embarked on the
analysis of these stars to see if this mechanism of a resonant interaction is
shared by other (most?) semiregular stars.  Preliminary results for stars such
as R UMi are very encouraging and again indicate a low dimensional chaotic
nature of the pulsations.  We also note that similar results have been obtained
with high quality data (Buchler, Koll\'ath and Cadmus 2002).  However, because
our goal is to extract quantitative information, such as dimensions, Lyapunov
exponents, etc., at this stage we feel that more work is necessary to establish
the robustness of the results that we have obtained.
 
We wish to thank the organizers for their generous support which made our
participation possible.
Our thanks also go to the AAVSO, AFOEV, BAAVSS and VSOLJ for allowing us
to use their data.  This work has been supported by NSF (AST9819608) and OTKA
(T038440).

\end{document}